\documentclass[prl,superscriptaddress,longbibliography,nofootinbib]{revtex4-2}

\usepackage[separate-uncertainty=true]{siunitx}
\usepackage{graphicx}
\usepackage{dcolumn}
\usepackage{bm}
\usepackage{hyperref}
\usepackage{float}
\usepackage{amssymb}
\usepackage{amsmath}
\usepackage{xcolor}
\usepackage[normalem]{ulem}

\begin{document}

\title{Quantum communication with itinerant surface acoustic wave phonons}

\author{\'E.~Dumur}
\altaffiliation[Present address: ]{Universit\'e Grenoble Alpes, CEA, INAC-Pheliqs, 38000 Grenoble, France}
\affiliation{Pritzker School of Molecular Engineering, University of Chicago, Chicago IL 60637, USA}
\affiliation{Argonne National Laboratory, Argonne IL 60439, USA}
\author{K.~J.~Satzinger}
\altaffiliation[Present address: ]{Google, Santa Barbara CA 93117, USA.}
\affiliation{Pritzker School of Molecular Engineering, University of Chicago, Chicago IL 60637, USA}
\affiliation{Department of Physics, University of California, Santa Barbara CA 93106, USA}
\author{G.~A.~Peairs}
\altaffiliation[Present address: ]{Amazon AWS, San Francisco CA 94105, USA.}
\affiliation{Pritzker School of Molecular Engineering, University of Chicago, Chicago IL 60637, USA}
\affiliation{Department of Physics, University of California, Santa Barbara CA 93106, USA}
\author{M.-H.~Chou}
\affiliation{Pritzker School of Molecular Engineering, University of Chicago, Chicago IL 60637, USA}
\affiliation{Department of Physics, University of Chicago, Chicago IL 60637, USA}
\author{A.~Bienfait}
\altaffiliation[Present address: ]{Universit\'e de Lyon, ENS de Lyon, Universit\'e Claude Bernard, CNRS, Laboratoire de Physique, F-69342 Lyon, France}
\affiliation{Pritzker School of Molecular Engineering, University of Chicago, Chicago IL 60637, USA}
\author{H.-S.~Chang}
\affiliation{Pritzker School of Molecular Engineering, University of Chicago, Chicago IL 60637, USA}
\author{C.~R.~Conner}
\affiliation{Pritzker School of Molecular Engineering, University of Chicago, Chicago IL 60637, USA}
\author{J.~Grebel}
\affiliation{Pritzker School of Molecular Engineering, University of Chicago, Chicago IL 60637, USA}
\author{R.~G.~Povey}
\affiliation{Pritzker School of Molecular Engineering, University of Chicago, Chicago IL 60637, USA}
\affiliation{Department of Physics, University of Chicago, Chicago IL 60637, USA}
\author{Y.~P.~Zhong}
\affiliation{Pritzker School of Molecular Engineering, University of Chicago, Chicago IL 60637, USA}
\author{A.~N.~Cleland}
\affiliation{Pritzker School of Molecular Engineering, University of Chicago, Chicago IL 60637, USA}
\affiliation{Argonne National Laboratory, Argonne IL 60439, USA}
\affiliation{Corresponding author: anc@uchicago.edu}

\date{\today}

\begin{abstract}
Surface acoustic waves are commonly used in classical electronics applications, and their use in quantum systems is beginning to be explored, as evidenced by recent experiments using acoustic Fabry-P\'erot resonators. Here we explore their use for quantum communication, where we demonstrate a single-phonon surface acoustic wave transmission line, which links two physically-separated qubit nodes. Each node comprises a microwave phonon transducer, an externally-controlled superconducting variable coupler, and a superconducting qubit. Using this system, precisely-shaped individual itinerant phonons are used to coherently transfer quantum information between the two physically-distinct quantum nodes, enabling the high-fidelity node-to-node transfer of quantum states as well as the generation of a two-node Bell state. We further explore the dispersive interactions between an itinerant phonon emitted from one node and interacting with the superconducting qubit in the remote node. The observed interactions between the phonon and the remote qubit promise future quantum optics-style experiments with itinerant phonons.
\end{abstract}

\maketitle

\section{Introduction}

Quantum communication is of significant interest for the generation of remote entanglement and the secure transmission of information, as well as for distributed quantum computing  \cite{bennett1993, 1997Cirac, bouwmeester1997, gottesman1999, dlcz2001, jiang2007, kimble2008}.
There are several demonstrations of long-distance quantum communication protocols using optical methods, in parallel with demonstrations of similar protocols using microwave-frequency photons, including Bell state entanglement of remote qubits as well as the transmission of multi-qubit entangled states \cite{pan2020, pan2020RMP, Chang2020, Kurpiers2018, Axline2018, Campagne-Ibarcq2018, Leung2019, Zhong2019, Zhong2021}.
Microwave-frequency phonons, as opposed to photons, can also be used for quantum communication as well as for coupling hybrid quantum systems \cite{vainsencher2016, Whiteley2019, peairs2020, mirh2020}, in the latter case taking advantage of the strong strain coupling in some optical as well as atomic-scale systems. Microwave-frequency acoustic resonators may be able to serve as very long-lived quantum memories \cite{maccabe2020}. Quantum communication protocols implemented with phonons are thus of significant scientific as well as practical interest. Recent advances in the quantum control of phonons include the creation and measurement of stationary phonon quantum states \cite{Chu2017, Chu2018, Satzinger2018}, the emission and absorption of phonons in an acoustic resonator \cite{Bienfait2019}, and the generation of entangled phonons in a phonon-mediated quantum eraser experiment \cite{Bienfait2020}.

Here we report the experimental realization of a phonon-based quantum communication channel, enabling the communication of quantum states via traveling phonons linking two physically-distinct quantum nodes. The phonons are emitted in the communication channel as short-duration acoustic pulses, sufficiently brief that the extent of the acoustic pulses is significantly less than the length of the channel, such that the phonons travel in a particle-like fashion along the channel, which we term itinerant.

\begin{figure*}[!ht]
    \centering
    \includegraphics[width=0.7\textwidth]{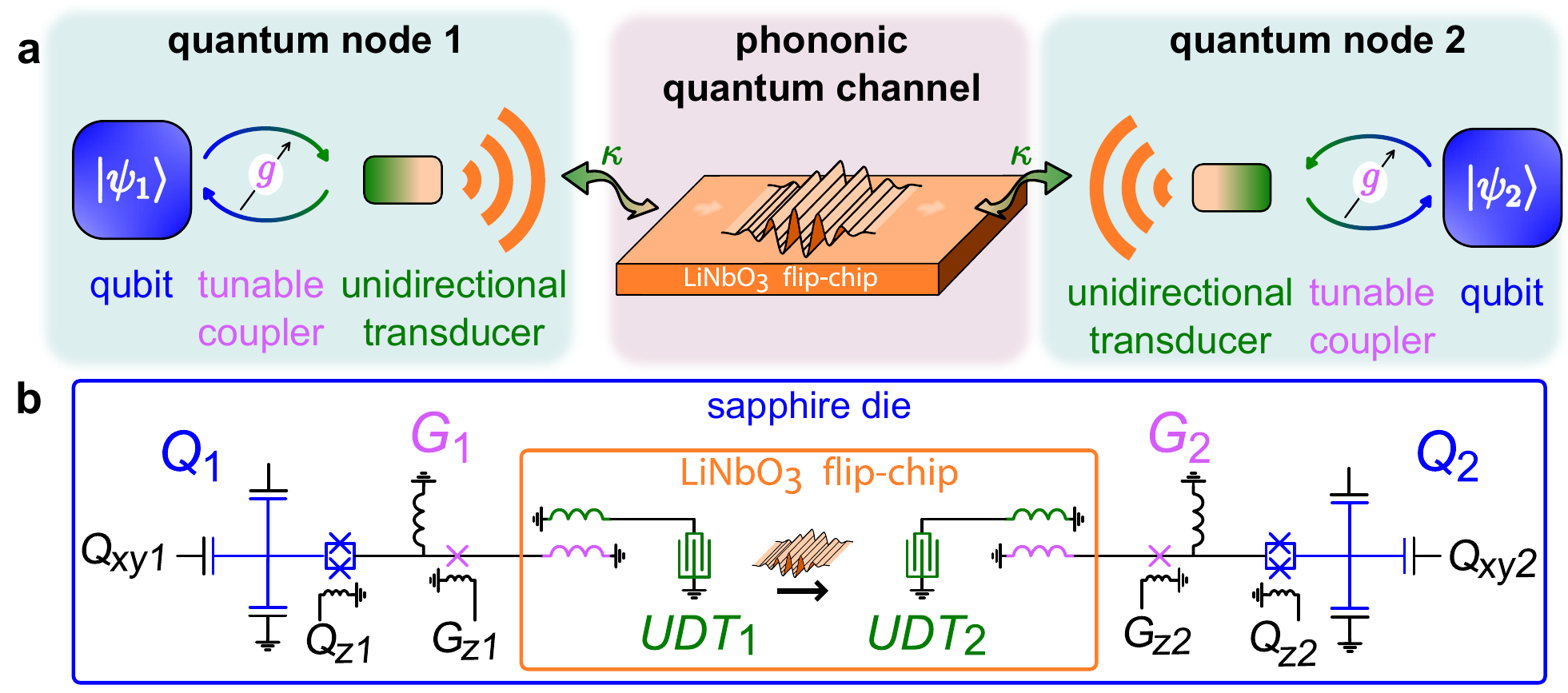}
    \caption{ Quantum delay line.
    \textbf{a} Schematic representation of the two quantum communication nodes and the phononic quantum channel. Each node comprises a superconducting qubit, a tunable coupler that allows shaping phonon release and capture, and a unidirectional phonon transducer. Details in the Methods section.
    \textbf{b} Circuit diagram of the assembled device.
    Each qubit $Q_j$ is excited through its dedicated $Q_{xy~j}$ microwave line and frequency-controlled through a separate $Q_{z\, j}$ flux-bias line, and each tunable coupler $G_k$ is controlled via its associated $G_{z\, k}$ flux-bias line.}
    \label{fig-intro}
\end{figure*}

The experimental system is shown schematically in Fig.~\ref{fig-intro}, with the physical setup in Fig.~\ref{fig-intro}\textbf{a} and the circuit schematic in Fig.~\ref{fig-intro}\textbf{b}. The \SI{2}{\milli\meter}-long phonon communication channel (\SI{500}{\nano\second} single-trip time) is terminated at each end by a specially-designed unidirectional interdigitated transducer, which is in turn connected to a superconducting qubit via a superconducting tunable coupler. The unidirectional transducers (UDTs) differ from conventional acoustic transducers, here emitting itinerant phonons in only one direction, as opposed to more standard bi-directional transducers, which emit excitations equally in two opposing directions (see Supplementary Note 1; a related but distinct design appears in \cite{2019Dumur}). We note this device differs from the experimental construction in e.g. Ref. \cite{Bienfait2019}, which uses a single bidirectional transducer in a Fabry-P\'erot cavity. In that experiment, a single phonon comprises acoustic excitations that travel in two opposing directions to distant acoustic mirrors, from which the excitations reflect and return to interfere constructively at the emitting transducer, where the excitation can be intercepted by one of two qubits. In the experiment here, two distinct unidirectional transducers are used to link two physically separate nodes. Each transducer is constructed to emit an acoustic excitation in only one direction, creating a significantly more flexible and general-purpose design, with physically separate and distinct phonon emitter and receiver.

We use this device to demonstrate two-node quantum state transfers as well as the phonon-mediated deterministic generation of an entangled Bell state, representing a significant advance over prior work, in which a single transducer was coupled to a Fabry-P\'erot acoustic cavity formed by two acoustic mirrors \cite{Chu2017, Chu2018, Satzinger2018, Bienfait2019, Bienfait2020}.
We also realize a single-phonon interferometer, using one qubit to emit and detect a traveling phonon, where the phonon is used to probe the state of the second qubit, effectively demonstrating the dispersive interaction of a photon (localized in the remote qubit) and a traveling phonon.
Finally, we demonstrate a Ramsey interferometer, using the second qubit to detect the presence of a traveling phonon emitted by the first qubit, thus interchanging the roles of the qubits in the previous experiment and demonstrating the versatility of this architecture.

\section*{RESULTS}

\subsection*{Phonon-mediated quantum state transfer}

\begin{figure}[!ht]
    \centering
    \includegraphics[width=0.4\textwidth]{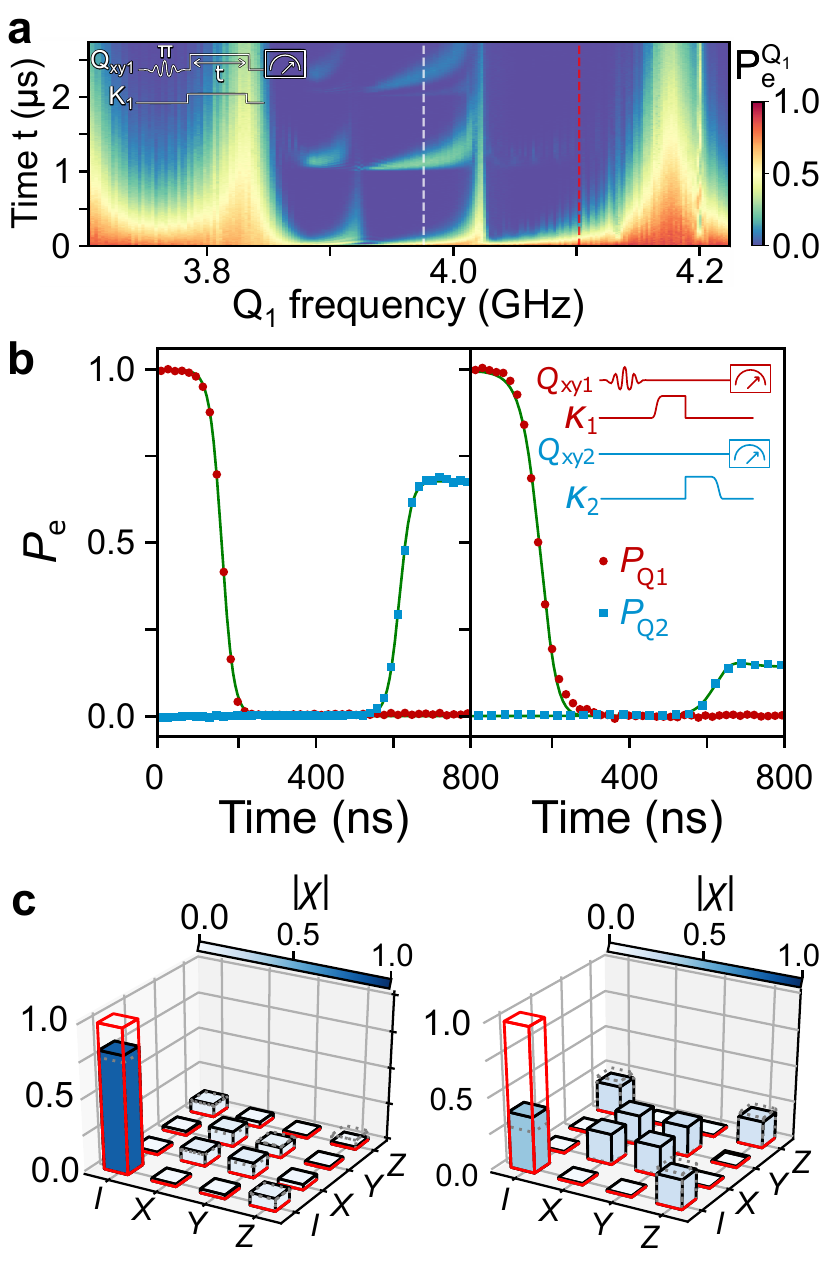}
    \caption{Phonon-mediated quantum state transfer and process tomography.
    \textbf{a} Measured $Q_1$ excited state population $P_e^{Q_1}$ as a function of time and $Q_1$ bare frequency, with coupler $G_1$ at an intermediate coupling $\kappa_1/2\pi = \SI{2.4}{\mega\hertz}$ (measured at \SI{3.976}{\giga\hertz}) and $G_2$ set to zero coupling. In this configuration, $Q_1$'s energy relaxation is dominated by phonon emission via $\mathrm{UDT}_{1}$, followed by traveling phonon dynamics. The white and red dashed lines indicate the unidirectional and bidirectional working frequencies, respectively (see text); inset shows the qubit excitation and measurement pulse sequence.
    \textbf{b} Quantum state transfer via a traveling phonon at the unidirectional (left) and bidirectional (right) operating frequencies. $Q_2$'s final population is $4.5$ times smaller for the bidirectional transfer compared to the unidirectional transfer, in line with simulations. Green solid lines are from a master equation simulation. Inset: Pulse sequence. For either process, $Q_1$'s emission rate is set to $\kappa_c^{\mathrm{uni|bi}}/2\pi = 10|\SI{6}{\mega\hertz}$, corresponding to a $81|\SI{138}{\nano\second}$ full-width-at-half-maximum (FWHM) phonon wavepacket.
    \textbf{c} Quantum process tomography for the unidirectional and bidirectional regimes, with process fidelities of $\mathcal{F}_\mathrm{uni} = \mathrm{Tr}~(\chi_\mathrm{exp} \cdot \chi_\mathrm{ideal}) = 82 \pm \SI{0.3}{\percent}$ and $\mathcal{F}_\mathrm{bi} = 39 \pm \SI{0.3}{\percent}$, respectively. Red solid lines show values expected for an ideal transfer; black dashed lines show master equation simulations, taking into account finite qubit coherence and phonon channel losses. Uncertainties are standard deviations from the mean.}
\label{fig-transfer}
\end{figure}

We first probe the interaction between the qubits and the phonon channel, as shown in Fig.~\ref{fig-transfer}\textbf{a}.
We excite $Q_1$ with a $\pi$ pulse, then set its coupler $G_1$ to an intermediate coupling, sufficient that $Q_1$'s relaxation is dominated by phonon emission.
We set $Q_2$'s coupler $G_2$ off during this measurement, so that $Q_2$ does not interact with the traveling phonon.
For frequencies inside the transducer's active band, from 3.87 to 4.01 GHz, where the emission is almost entirely unidirectional itinerant phonons, we observe a time-delayed revival of qubit $Q_1$'s excited state population $P_e^{Q_1}$ at times that are multiples of the phonon round-trip time $\tau_\mathrm{RT} \sim \SI{1}{\micro\second}$, each revival corresponding to the traveling phonon reflecting off the other transducer before re-exciting $Q_1$.
Outside the unidirectional band, we see a complex structure in  $P_e$ as a function of frequency and interaction time, with broad swings of width $\sim \SI{150}{\mega\hertz}$ superposed with narrow oscillations of width $\sim \SI{7}{\mega\hertz}$.
The broad swings and finer details are in accordance with expectations (see Supplementary Note 1) \cite{Morg2007}.

The itinerant phonon experiments are performed at the working frequency $\omega_{Q_{1,2}}^\mathrm{uni}/2\pi= \SI{3.976}{\giga\hertz}$, inside the unidirectional band.
By working outside this band, we can explore the regime where the transducers are effectively bidirectional, using the second working frequency $\omega_{Q_{1,2}}^\mathrm{bi}/2\pi = \SI{4.102}{\giga\hertz}$.
These frequencies are marked by the dashed white and red lines, respectively, in Fig.~\ref{fig-transfer}\textbf{a}.

To maximize the efficiency of phonon-mediated quantum state transfers, we need to carefully shape the emission and absorption of the phonon wave packet, which is done by time-dependent control of the coupling between the qubit and its transducer \cite{2011Korotkov,Zhong2019,Kurpiers2018,Axline2018,Campagne-Ibarcq2018,Leung2019,Bienfait2019}.
We experimentally optimize the transfer efficiency, with results shown in Fig.~\ref{fig-transfer}\textbf{b} for both the unidirectional (left) and bidirectional (right) regimes.
The transfer starts with the shaped emission of a phonon, shown by the decrease of $Q_1$'s excited state population with the expected time dependence.
Both qubits then remain in their ground states until the phonon reaches $Q_2$, which absorbs the itinerant phonon, following the expected time dependence, and ultimately reaching a plateau once the transfer is complete.
The total transfer takes $\sim \SI{700}{\nano\second}$, including the $\sim \SI{500}{\nano\second}$ phonon travel time.
The final $Q_2$ population reaches a maximum of \SI{68}{\percent} for the unidirectional transfer, limited mostly by phonon loss in the channel.
For the bidirectional transfer, the final $Q_2$ population reaches \SI{15}{\percent}, $4.5$ times less than the unidirectional population, which is \SI{12}{\percent} higher than the ideal value, demonstrating good agreement with theory and excellent unidirectionality for the transducer design.
We simulate the transfer process using a cascaded quantum input-output model \cite{Bienfait2019} (solid green line).
From this model we estimate that phonon loss reduces the final unidirectional transfer efficiency by \SI{27}{\percent}, and the finite $Q_1$ and $Q_2$ coherence times reduce the fidelity by \SI{1}{\percent} and \SI{2}{\percent}, respectively.
We note that an equivalent photon travel time would require a $\sim \SI{100}{\meter}$ long coaxial cable, illustrating the very long delays achievable with phonon-based quantum channels.

In Fig.~\ref{fig-transfer}\textbf{c}, we show quantum process tomography for both regimes.
For the unidirectional process, we find a process fidelity of $\mathcal{F}^\mathrm{uni} = \SI{82\pm0.3}{\percent}$, while for the bidirectional regime, the process fidelity is limited to $\mathcal{F}^\mathrm{bi} = \SI{39\pm0.3}{\percent}$.
We compare these experimental process fidelities with predictions, and find trace distances $d = \sqrt{\mathrm{Tr}~(\chi_\mathrm{exp} - \chi_\mathrm{sim})^2} = 0.07$ and $0.3$ for the unidirectional and bidirectional regimes. The contrast in fidelities and trace distances underlines the importance of the unidirectional transducers.

\subsection*{Traveling phonon-mediated remote entanglement}

\begin{figure}[!ht]
    \centering
    \includegraphics[width=0.4\textwidth]{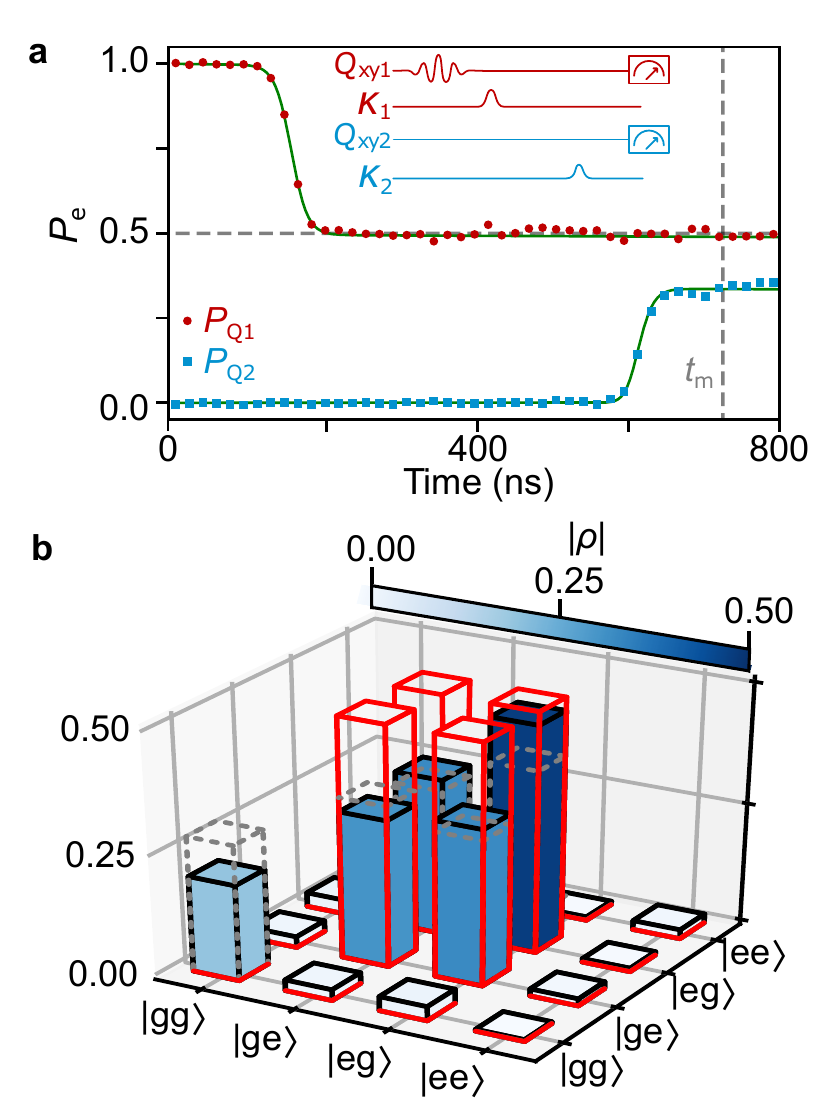}
    \caption{Phonon-mediated Bell state generation.
    \textbf{a} Inset: Pulse sequence for Bell state generation. Main panel: Excited state probabilities for $Q_1$ (red) and $Q_2$ (blue) as a function of time. Green lines are results from a master equation simulation. The final state is analyzed at $t_m = \SI{725}{ns}$ (gray dashed line).
    \textbf{b} Bell state density matrix, absolute values, without readout correction, measured at $t_\mathrm{m}$. Red solid lines show values expected for an ideal Bell state; black dashed lines show simulation results including qubit coherence and phonon channel losses.}
    \label{fig-bell}
\end{figure}

We further explore the capabilities of itinerant phonon communication by performing a phonon-mediated remote entanglement of the two qubits, shown in Fig.~\ref{fig-bell}. The protocol is similar to that for the quantum state transfer, except here we calibrate the emission pulse to only emit $Q_1$'s excitation as a phonon with a probability of 1/2, meaning that immediately following the `half-emission,' with qubit $Q_2$ in the ground state, the system is ideally in the state $(|e0g\rangle+|g1g\rangle)/\sqrt{2}$ (writing the state $|Q_1 \, \gamma \, Q_2\rangle$ where $\gamma$ represents the itinerant phonon). During the time the emitted `half-phonon' travels along the phonon channel, $Q_1$'s remaining excitation decays following $Q_1$'s intrinsic $T_1$ time, with $Q_1$'s coupling to the channel set to zero. The traveling half-phonon is then captured by $Q_2$, generating a Bell state $|\psi\rangle = (|eg\rangle + e^{\mathrm{i} \varphi}|ge\rangle)/\sqrt{2}$ between the two qubits, with $\varphi$ a relative phase.

Figure~\ref{fig-bell}\textbf{a} shows the time-dependent qubit state populations $P_e$ for each qubit, which agree well with a master equation simulation.
Following capture of the half-phonon, we perform quantum state tomography at time $t_\mathrm{m} = \SI{750}{\nano\second}$; these measurements are used to reconstruct the two-qubit density matrix $\rho$ shown in Fig.~\ref{fig-bell}\textbf{b}.
We find a Bell state fidelity $\mathcal{F}_\mathrm{Bell} = \mathrm{Tr}~(\rho_\mathrm{ideal} \cdot \rho) = \SI{72}{\percent}$ and a concurrence $\mathcal{C}=\SI{0.53}{}$, close to the master equation simulation results, with a trace distance $d^\mathrm{Bell} = \sqrt{\mathrm{Tr}~(\rho_\mathrm{exp} - \rho_\mathrm{sim})^2} = \SI{0.13}{}$.

\subsection*{Phonon-qubit dispersive interaction}

\begin{figure}[!ht]
    \centering
    \includegraphics[width=0.8\textwidth]{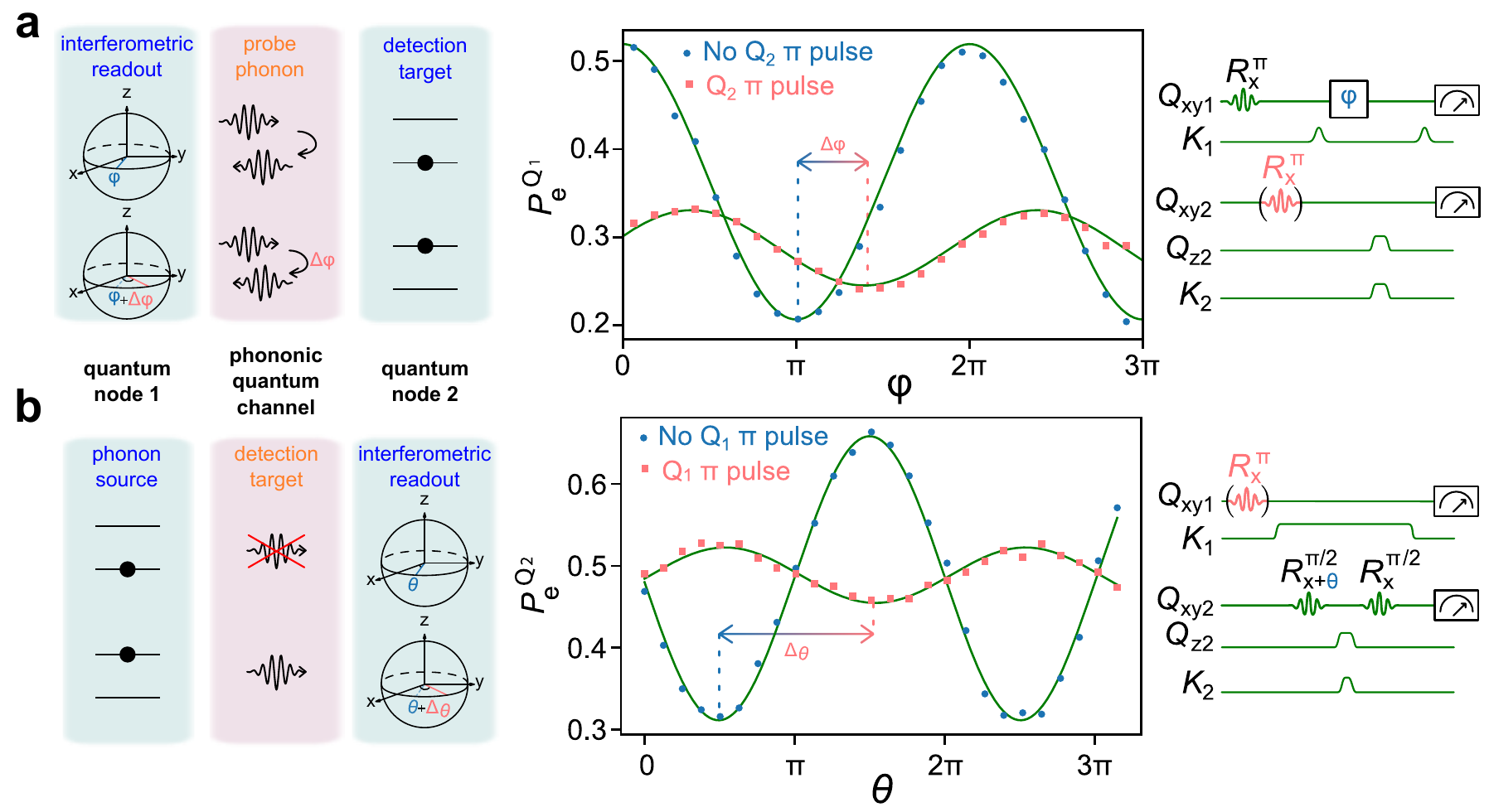}
    \caption{Phonon interferomeric probe.
    Dispersive state-dependent interferometric probe. \textbf{a} Schematic roles played by each element. \textbf{b}  Blue (salmon) points show the $\varphi$ dependence of $Q_1$'s final $P_e$ when $Q_2$ is in $|g\rangle$ ($|e\rangle$), showing a dependence of the phonon phase on $Q_2$'s state, with a shift of $\sim 0.4 \pi$~rad. \textbf{c} Qubit pulse sequences (see text for details).
    \textbf{d} Schematic roles played by each element for the dispersive phase-dependent interferometric probe.  \textbf{e} Ramsey interference in $Q_2$ (blue circles) reveals $Q_2$'s dependence on the relative phase with the phonon. When repeating the same protocol with no initial $\pi$ pulse on $Q_1$, we measure the Ramsey interference of $Q_2$ (salmon squares), shifted by $\pi$ compared to the first measurement. Blue (salmon) points show the $\varphi$ dependence of $Q_1$'s final $P_e$ when $Q_2$ is in $|g\rangle$ ($|e\rangle$), showing a dependence of the phonon phase on $Q_2$'s state, with a shift of $\sim 0.4 \pi$~rad. \textbf{f}  Qubit pulse sequences, similar to \textbf{c} except $Q_2$ is always placed in $(|g\rangle + e^{\mathrm{i} \theta} |e\rangle)/\sqrt{2}$ and the $\theta$-dependent $P_e^{Q2}$ is measured. }
\label{fig.phonon}
\end{figure}

Sensing traveling phonons without absorbing them would provide a highly useful capability, as would being able to use a traveling phonon as a probe of a remote quantum system, which we explore in a pair of related experiments.
First, we use a traveling phonon as a probe of a remote quantum two-level system, shown in Fig.~\ref{fig.phonon}\textbf{a}.
We use qubit $Q_1$ as the emitter and receiver of a ``half-phonon'' that is detected interferometrically \cite{Bienfait2019,Bienfait2020} when returning to $Q_1$.
This allows us to measure how the phase of the traveling phonon is affected by interacting dispersively with qubit $Q_2$, which serves as a stand-in for a generic quantum system.

The pulse sequence for this state detection is shown to the right in Fig.~\ref{fig.phonon}\textbf{a}: We first prepare $Q_1$ in its excited state, and emit a half-phonon, which reflects from the distant transducer, whose coupling to $Q_2$ is turned on during the reflection process, and the half-phonon interacts with $Q_1$ on its return.
During the half-phonon transit, we briefly shift $Q_1$'s frequency so that $Q_1$'s excited state acquires a relative phase $\varphi$, yielding an interferometric interaction with the returning half-phonon, either interfering constructively to return $Q_1$ towards its excited state, or destructively and having $Q_1$ emit its remaining energy and relax to its ground state.
In Fig.~\ref{fig.phonon}\textbf{a}, we show the final $Q_1$ population as a function of the phase $\varphi$ (blue points), showing a characteristic interference pattern with a visibility of \SI{32}{\percent}.

We repeat the experiment with $Q_2$ excited by a $\pi$ pulse at the beginning of the experiment, with experiment otherwise unchanged; the results are shown in Fig.~\ref{fig.phonon}\textbf{a} (salmon points).
There are three effects on the oscillation pattern: A slight increase in the oscillation minima, attributed to a decrease of the phonon coherence \cite{Bienfait2019} in its interaction with $Q_2$; a more marked reduction of visibility attributed to an inadequate absorption of the phonon wave packet; and, most significantly, a phase shift of $\Delta \varphi_\mathrm{exp}=0.40 \pi$ attributed to the dispersive interaction between $Q_2$ and the traveling half-phonon, close to our fit-free simulated value of $\Delta \varphi_\mathrm{sim}=0.41 \pi$ (see Supplementary Note 9).
This last effect points to the interesting possibility of using phonons as dispersive probes of other quantum systems.

In a separate experiment, shown in Fig.~\ref{fig.phonon}\textbf{b}, we swap the roles of the qubits, so $Q_2$ is now used as a dispersive probe for the phonon released by $Q_1$, using a Ramsey fringe measurement of $Q_2$.
The pulse sequence is shown to the right in Fig.~\ref{fig.phonon}\textbf{b}, where $Q_2$ is placed in the state $(|g\rangle + e^{\mathrm{i} \theta} |e\rangle)/\sqrt{2}$ by the initial $\pi/2$ rotation, performed about an axis rotated in the $x-y$ plane of the Bloch sphere by $\theta$, and the $\theta$-dependent evolution of $Q_2$ is compared for where $Q_1$ is not excited (no probe phonon) to where $Q_1$ is excited and $Q_2$ interacts with the subsequently-released traveling phonon.
In the latter case the Ramsey fringe visibility is reduced, which we attribute to leakage from $Q_2$ into the phononic channel, but we again observe a significant phase shift, here as high as $\Delta \theta_\mathrm{exp} = 0.95 \, \pi$ close to our simulation $\Delta \theta_\mathrm{sim} = 0.99 \, \pi$.

\section*{DISCUSSION}

In conclusion, we demonstrate controlled phonon-mediated quantum state transfer and remote entanglement between two quantum nodes, each node comprising a superconducting qubit with a time-variable coupler, using individual itinerant SAW phonons traveling in an acoustic transmission line after a controlled, on-demand release, followed by capture.
Using this architecture, we also demonstrate the dispersive interaction between an itinerant phonon and a superconducting qubit.
These results have been made possible by the integration of broadband, highly unidirectional transducer in a \SI{2}{\milli\meter} long phonon communication channel, as well as the use of a quantum state protocol requiring tunable coupling to each qubit node \cite{1997Cirac}.
Achieving a quite impressive quantum state transfer fidelity of \SI{82 \pm 0.2}{\percent}, limited by the loss in the phonon channel, this platform paves the way for quantum-optics-like experiments realized with individual phonons instead of photons.

\section*{METHODS}
\subsection{Device fabrication and characterization}
The device used in these experiments comprises two dies, a sapphire die with the two superconducting qubits ($Q_1$ and $Q_2$), and their associated tunable couplers ($G_1$ and $G_2$, respectively), as well as control and readout wiring, and a lithium niobate die with the phononic channel and the two unidirectional transducers. The two dies are fabricated separately then flip-chip assembled \cite{Satzinger2019}.
The full circuit schematic is shown in Fig.~\ref{fig-intro}\textbf{b},

The acoustic die is fabricated using a single layer of $\sim \SI{25}{\nano\meter}$ thick aluminium patterned by PMMA liftoff on a LiNbO$_3$ wafer, \SI{500}{\micro\meter} thick. The central part of the acoustic device is the $\ell = \SI{2}{\milli\meter}$-long phononic channel, with width $W = \SI{150}{\micro\meter}$, terminated at each end by a unidirectional transducer (UDT$_{1,2}$).

The unidirectional transducers (UDTs) are described more completely in Supplementary Note 1. Briefly, the two (nominally identical) unidirectional transducers (UDTs) each comprise a standard bi-directional interdigitated transducer (IDT) combined with an acoustic mirror (a reflective grating). The IDT emits equal-amplitude acoustic excitations in opposite directions, one towards and the other away from the second UDT. The acoustic mirror, placed immediately adjacent to the IDT on the side opposite the second UDT, reflects its incident excitation back towards the second UDT, such that it interferes constructively with the other excitation. Each UDT is coupled inductively to one of the two qubits.

We have separately characterized similar IDT-mirror designs, where in the frequency band from about $3.85$ to $\SI{4}{\giga\hertz}$, excellent directionality is achieved, with emission from the UDT almost entirely directed away from the mirror. Typical directivities are greater than $\SI{20}{\decibel}$. Outside this unidirectional band, the mirrors are less effective and the devices emit more strongly in both directions \cite{2019Dumur}.

The superconducting qubit die is fabricated on \SI{430}{\micro\meter}-thick sapphire using standard lithographic processing \cite{Zhong2019}. The qubits $Q_{1,2}$ are tunable xmon-style qubits \cite{2007Koch,2013Barends}, where each qubit's frequency is controlled by a flux line $Q_\mathrm{z1,z2}$, and excited using a capacitively-coupled microwave line $Q_\mathrm{xy1,xy2}$. Each qubit is coupled to the SAW chip through a superconducting tunable coupler $G_{1,2}$, whose coupling is controlled \cite{2014Chen} using external flux lines $G_\mathrm{z1,z2}$. Qubit states are inferred from standard dispersive measurements using a separate readout resonator for each qubit. The readout resonators are connected to a common readout line; more details are given in the Supplementary Note 1.

The qubits are characterized with their couplers turned off (see Supplementary Notes 6 and 7). At the qubit idle frequency $\omega_\mathrm{idle}/(2\pi) \sim \SI{4.3}{\giga\hertz}$, we find the qubits have an energy relaxation time $T_1 = \SI{57}{\micro\second}$ ($Q_1$) and $\SI{38}{\micro\second}$ ($Q_2$), with a coherence time $T_2^\mathrm{Ramsey} = \SI{1.11}{\micro\second}$ ($Q_1$) and $\SI{0.88}{\micro\second}$ ($Q_2$) (most likely limited by flux noise as the qubits are tuned far away from their flux-insensitive point). These times demonstrate the potential for excellent qubit coherence when using a flip-chip assembly \cite{Satzinger2019}.

\section*{DATA AVAILABILITY}
The data that support the findings of this study are available from the corresponding author upon reasonable request. Correspondence and requests for materials should be addressed to A. N. Cleland (anc@uchicago.edu).

\section*{ACKNOWLEDGEMENTS}
Devices and experiments were supported by the Air Force Office of Scientific Research and the Army Research Laboratory, and material for this work was supported by the Department of Energy (DOE).
\'E.D. was supported by LDRD funds from Argonne National Laboratory, K.J.S. was supported by NSF GRFP (NSF DGE-1144085) and A.N.C. was supported by the DOE, Office of Basic Energy Sciences.
This work was partially supported by the UChicago MRSEC (NSF DMR-2011854), AFOSR under award FA9550-20-1-0270, the NSF QLCI for HQAN (NSF Award 2016136) and made use of the Pritzker Nanofabrication Facility, which receives support from SHyNE, a node of the National Science Foundation's National Nanotechnology Coordinated Infrastructure (NSF NNCI ECCS-2025633).

\section*{COMPETING INTERESTS}
The authors declare no competing financial or non-financial interests.

\section*{AUTHOR CONTRIBUTIONS}
{\'E}.D. designed, and fabricated the devices.
{\'E}.D. performed the experiment and analyzed the data.
K.J.S., G.A.P. and M.-H.C. participated to the design process of the unidirectional transducer.
{\'E}.D., K.J.S., A.B., H.-S.C., J.G., Y.P.Z. developed the fabrication process of the superconducting circuit.
{\'E}.D., K.J.S. and A.B. wrote code to model surface acoustic wave.
A.N.C. advised on all efforts.
All authors contributed to discussion and production of the manuscript.

\clearpage

\begin{center}
\textbf{\Large{Supplementary Discussion for ``Quantum communication with itinerant surface acoustic wave phonons''}}
\end{center}

\section{Device details and new unidirectional transducer design}

\begin{figure}[!ht]
\centering
\includegraphics[width=0.8\textwidth]{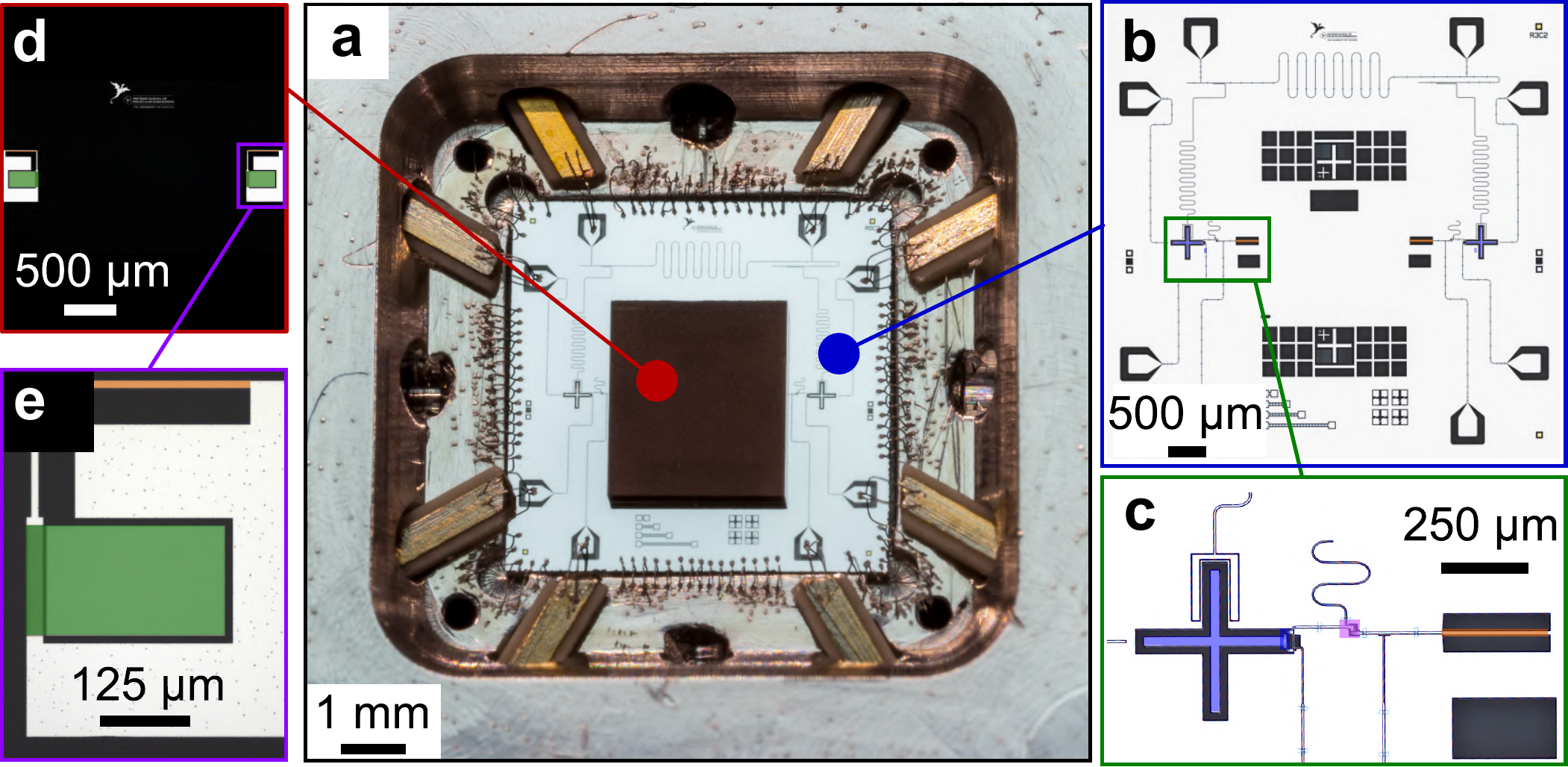}
\caption{Device construction.
(a) Micrograph of the assembled flip-chip device bonded in its sample holder.
(b, c) Optical micrograph of the sapphire qubit die with the two qubits ($Q_{1,2}$, blue), connected to their tunable couplers ($G_{1,2}$, purple).
The sapphire die is electrically coupled to the lithium niobate flip-chip die through overlapping inductive leads (orange).
(d, e) Optical micrograph of the lithium niobate die with two unidirectional transducers ($\mathrm{UDT}_{1,2}$, green).}
\label{fig.device}
\end{figure}

The acoustic die is fabricated using a single layer of $\sim \SI{25}{\nano\meter}$ thick aluminium patterned by PMMA liftoff on a LiNbO$_3$ wafer, \SI{500}{\micro\meter} thick. The central part of the acoustic device is the $\ell = \SI{2}{\milli\meter}$-long phononic channel, with width $W = \SI{150}{\micro\meter}$, terminated at each end by a novel unidirectional transducer (UDT$_{1,2}$), see Fig.~\ref{fig.device}.

The two (nominally identical) unidirectional transducers (UDTs) each comprise two main components: A standard bi-directional interdigitated transducer (IDT) combined with an acoustic mirror. The IDT emits equal-amplitude acoustic excitations in opposite directions, one towards ($T$) and the other away ($A$) from the second UDT. The acoustic mirror, placed immediately adjacent to the IDT on the side opposite the second UDT, reflects its incident excitation $A$ back towards the second UDT, such that it interferes constructively with the excitation $T$.

The IDT itself comprises a double-electrode tapered IDT \cite{Morg2007} with $N^\mathrm{IDT}=24$ cells, designed for a center wavelength $\lambda_0^\mathrm{IDT} = \SI{975}{\nano\meter}$. The acoustic mirror comprises a tapered open-circuit reflective grating \cite{Morg2007} with $N^\mathrm{mirror} = 488$ electrodes, placed \SI{500}{\nano\meter} from the last finger of the IDT, with a design center wavelength $\lambda_0^\mathrm{mirror} = \SI{1}{\micro\meter}$. The grating center wavelength is slightly larger than the IDT center wavelength to compensate for frequency-loading of the IDT by the superconducting circuit. Each IDT is inductively coupled to one of the qubits on the sapphire chip.

We have separately characterized similar IDT-mirror designs, where in the frequency band from about $3.85$ to $\SI{4}{\giga\hertz}$, excellent directionality is achieved, with emission from the UDT almost entirely directed away from the mirror. Typical directivities are greater than $\SI{20}{\decibel}$. Outside this unidirectional band, the mirrors are less effective and the devices emit more strongly in both directions \cite{2019Dumur}.

As mentioned in the main text, from measurements of the response of a single qubit coupled to its transducer, shown in Fig.~1 in the main text, we estimate the unidirectional band of the transducers to be about \SI{140}{\mega\hertz} wide, centered at \SI{3.95}{\giga\hertz}, in accordance with room-temperature measurements of similar devices. Outside the unidirectional active band, we see a complex structure in  $P_e$ as a function of frequency and interaction time, with broad swings of width $\sim \SI{150}{\mega\hertz}$ superposed with narrow oscillations of width $\sim \SI{7}{\mega\hertz}$. The broad swings and finer details are in accordance with expectations, including simulations in which we construct the detailed response of the transducers and the channel using $P$-matrix theory \cite{Morg2007}.

We note that inside the active unidirectional transducer band, a small feature appears at \SI{3.92}{\giga\hertz}, which we interpret as a destructive interference between the transducer and its adjacent mirror, which impedes phonon emission over a bandwidth of $\sim \SI{15}{\mega\hertz}$. This allows us to extract the effective distance between the transducer and the mirror $d_\mathrm{IDT-mirror} = \SI{140}{\nano\meter}$, which is larger than the design value of \SI{50}{\nano\meter}, perhaps due to an under-compensation of the effect of the transducer and mirror tapers. This interference however has no effect on measurements performed away from this frequency.

The superconducting qubit die is fabricated on \SI{430}{\micro\meter}-thick sapphire using standard lithographic processing \cite{Zhong2019}.
The qubits $Q_{1,2}$ are tunable xmon-style qubits \cite{2007Koch,2013Barends}, where each qubit's frequency is controlled by a flux line $Q_\mathrm{z1,z2}$, and excited using a capacitively-coupled microwave line $Q_\mathrm{xy1,xy2}$.
Each qubit is coupled to the SAW chip through a superconducting tunable coupler $G_{1,2}$, whose coupling is controlled \cite{2014Chen} using external flux lines $G_\mathrm{z1,z2}$.
Qubit states are inferred from standard dispersive measurements using a separate readout resonator for each qubit. The readout resonators are connected to a common readout line.

\section{Flux crosstalk calibration}

All experiments rely on a precise control of the tunable coupler settings and the qubit frequencies.
Tuning is achieved by applying local flux bias using microwave cables connected to room temperature digital electronics.
We note that an appreciable level of cross-talk was measured between the each flux control line of our chip and the flux-tunable elements.
This crosstalk was calibrated by measuring the frequency shift of each qubit due to each flux lines.
We then assumed that the crosstalk was symmetric between each coupler-qubit pair, for example $Q_1$ to $G_1$, allowing us to obtain a crosstalk matrix which was used through linear inversion to cancel the crosstalk.

\section{Calibration of qubit frequency shifts due to tunable couplers}

The coupler design\cite{2014Chen} is an inductive $\pi$ network whose center inductance is a Josephson junction, whose phase and thus effective inductance can be tuned from a finite inductance to a very large positive inductance and then immediately to a large negative inductance and back to a finite (negative) inductance by applying an external flux.
This results in varying the coupling from a finite value through zero to a (negative) finite value.
Tuning the coupler however affects its associated qubit frequency, as the inductance of the coupler determines in part the qubit frequency.
We calibrate the flux required on the qubit flux bias line to cancel this induced frequency shift.
By doing this for both qubit-coupler pairs, we ensure control of the qubit frequency during any experiment.

\section{Phononic channel loss characterization}

To estimate the phononic channel loss, we must ensure there is no parallel loss from the qubits.
This is achieved by turning off the coupling between each qubit and its associated transducer.
With the coupler off, we measure the intrinsic qubit lifetime; see Table.~\ref{tab:supp-1}.
We next have to calibrate the qubits and couplers to achieve an optimal shaped phonon envelope.
This calibration is checked by monitoring the qubit excited state population time dependence $P_e(t)$.

Given the emitted phonon amplitude envelope $c_\mathrm{out} (t)$, the corresponding qubit excited state population $P_\mathrm{e} (t)$ can be determined, in the absence of other qubit loss processes.
Experimentally, we compare the measured $P_\mathrm{e}^\mathrm{m} (t)$ to the expected $P^\mathrm{e}_\mathrm{e} (t)$ for each $c_\mathrm{out} (t)$.
This process is relatively straightforward in the unidirectional regime but somewhat more challenging in the bidirectional regime, likely due to the slower emission rate in the less symmetrical admittance environment.

\begin{figure}[!ht]
\centering
\includegraphics[width=0.5\textwidth]{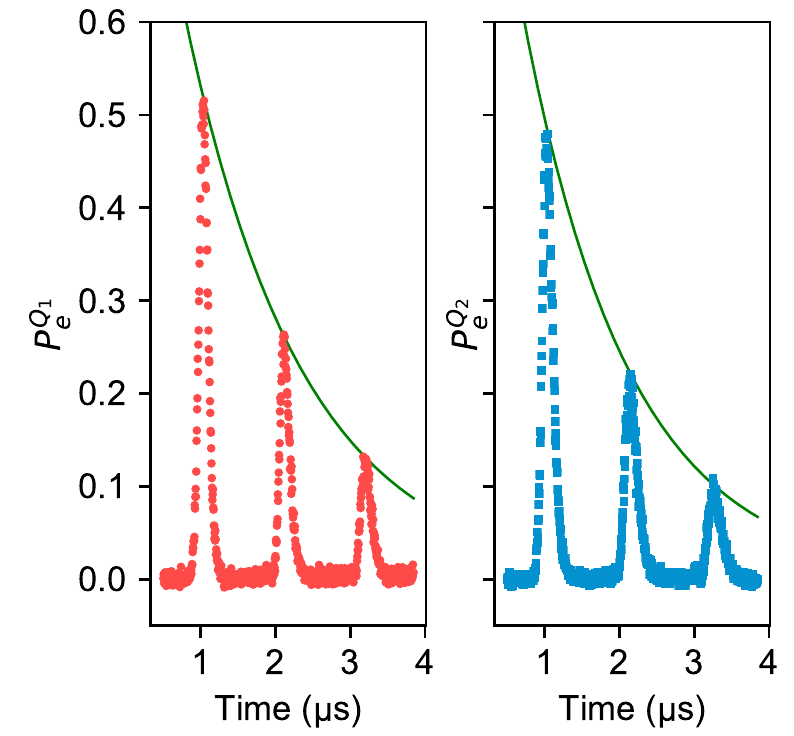}
\caption{Optimized multiple phonon emission and absorption from $Q_1$, left, and $Q_2$, right.
In this measurement the losses in the phononic channel are the dominant loss.
From a simple exponential fit we extract a mean $T_\mathrm{saw}=\SI{1.5}{\micro\second}$, which corresponds to a loss rate $\alpha=\SI{173}{\neper\per\meter}$ and a free space velocity $v_\mathrm{f}=\SI{3863}{\meter\per\second}$.}
\label{fig.supp-tsaw}
\end{figure}

We then measure the qubit excited state population while sweeping the delay between the emission and absorption; this is shown in Fig.~\ref{fig.supp-tsaw}.
We observe a mathematically geometrical damping of the qubit excited state revival as function of the number of round-trips.
From this measurement we extract a mechanical loss of $\alpha=\SI{173}{\neper\per\meter}$.

\section{Phonon thermalization}

\begin{figure}[!ht]
\centering
\includegraphics[width=0.5\textwidth]{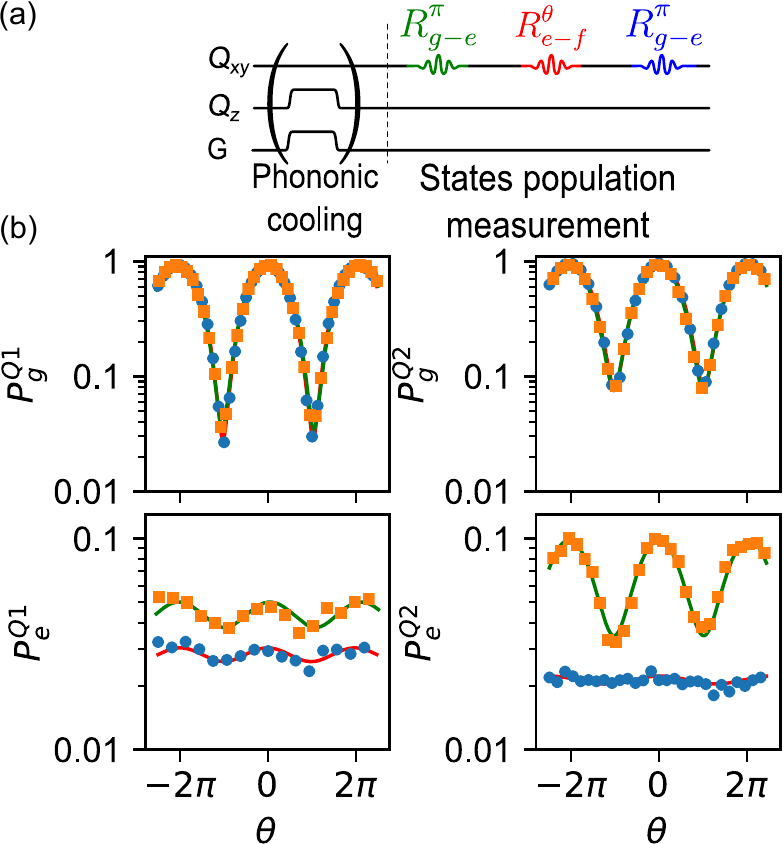}
\caption{Qubit cooling and thermometry.
(a) Pulse sequences showing the phonon-based cooling procedure followed by a qubit excited state measurement.
(b) Measurement of the qubit state populations $P_g$ and $P_e$, for $Q_1$ on the left and $Q_2$ on the right.
The upper (lower) panels show the population of the qubit ground (excited) state, respectively.
Blue circles and orange squares show the measurement results with and without phononic cooling.
Solid lines show simple cosine fits used to extract the relative amplitude of each curve, from which the qubit temperatures are deduced.}
\label{fig.thermometry}
\end{figure}

We typically find that our superconducting qubits have higher quiescent excited state populations than indicated by the environmental temperature, and higher than that of the phononic devices.
For every measurement reported in the main text and in the supplementary, we used a cooling procedure that takes advantage of the lower temperature phonon environment.
We set the qubit couplers to maximum coupling, and bring each qubit into resonance with their respective UDT at $\sim \SI{4.15}{\giga\hertz}$, outside of the unidirectional bandwidth.
The qubits are left in this state for $\sim \SI{300}{\nano\second}$, long enough for the qubits to relax to the environmental phononic temperature.

To measure the effect of this cooling protocol, we implement the experiment described in Ref.~\cite{PhysRevLett.110.120501} to measure the qubit excited state population, with and without the cooling procedure.
The pulse sequence and results are shown in Fig.~\ref{fig.thermometry}.
We measure that cooling procedure consistently lowers the qubit excited state population from \SI{1}{\percent} (\SI{7}{\percent}) to  \SI{0.5}{\percent} (\SI{0.7}{\percent}) for $Q_1$ ($Q_2$).

\section{Multiplexed qubit readout: Visibility and readout correction}

\begin{figure}[!ht]
\centering
\includegraphics[width=0.5\textwidth]{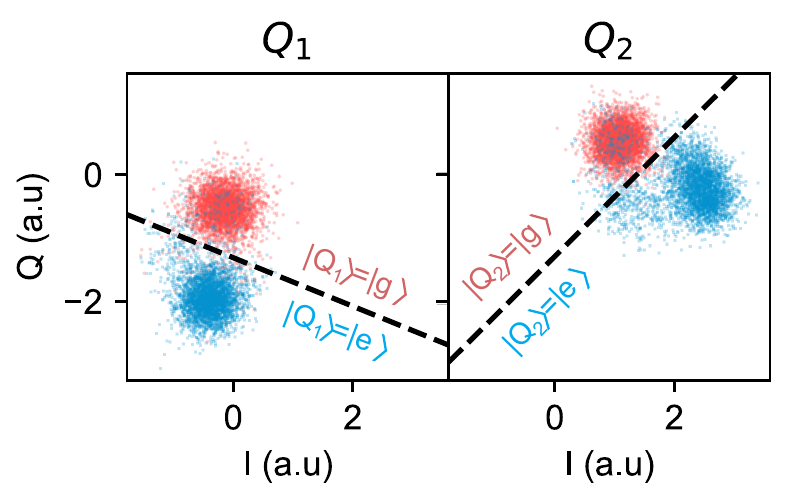}
\caption{Multi-qubit readout measurement.
Each measurement consists of placing the qubit in its ground or excited state, followed by a measurement of the amplitude and phase of the readout signal.
This figure shows the typical qubit ground and excited state measurement, after optimization of the control and readout sequences.}
\label{fig.supp-iq}
\end{figure}

The qubit ground, excited and third states ($|g\rangle, |e\rangle$ and $|f\rangle$) are discriminated using a dispersive readout scheme, where the frequency shift induced by each qubit $Q_{1, 2}$ on its dedicated readout resonator $R_{1,2}$ is used to infer the qubit state.
These are single-shot measurements that ideally leave the qubit in its measured state post-measurement.
Each readout resonator is inductively-coupled to a Purcell filter \cite{Reed2010a,Jeffrey2014} to allow for a stronger qubit-resonator coupling without degrading the qubit lifetime, allowing for a faster qubit state readout.
The Purcell filter consists of a short-circuited $\lambda/2$ coplanar waveguide resonator, with an under-coupled input port and an over-coupled output port, such that the ratio of the coupling quality factors $Q_\mathrm{in}/Q_\mathrm{out} > 100$ by design.
This maximizes the readout signal coupled into the common output transmission line that goes to the first-stage amplifier.

The output port is connected to a double junction isolator (Low Noise Factory) placed on the dilution refrigerator mixing chamber stage, followed by a high electron mobility transistor (HEMT) amplifier (Low Noise Factory) at the \SI{4}{\kelvin} stage.
Further amplification is provided by room temperature amplifiers (MITEQ Corp.).
Detection is performed using a heterodyne scheme implemented using custom electronics with a \SI{1}{\giga\hertz} sampling rate and a \SI{250}{\mega\hertz} bandwidth, which extracts the amplitude and phase, or equivalently the $I$ and $Q$ quadrature coordinates, of the readout signal.
Qubit state discrimination uses a threshold optimization technique shown in Fig.~\ref{fig.supp-iq}, where the $I-Q$ plane is divided into two parts and each measured $I-Q$ coordinate pair is assigned to the qubit ground $|g\rangle$ or excited $|e\rangle$ state.

Typical single qubit readout fidelities in this experiment were $\mathcal{F}_\mathrm{readout}^{Q_1} \sim \SI{95}{\percent}$ and  $\mathcal{F}_\mathrm{readout}^{Q_2} \sim \SI{94}{\percent}$.
We chose to measure both qubits simultaneously using multiplexed readout signals, with a single \SI{1}{\micro\second} readout pulse applied and measured through the readout line.
We characterize the readout visibility by measuring the matrix $V$ defined as $P_\mathrm{meas} = V P_\mathrm{exp}$, the transformation between the measured probability vector and the probability vector expected given the control sequence, in the two-qubit basis $\{|gg\rangle,|ge\rangle,|eg\rangle,|ee\rangle\}$.
Typical readout characterizations are averaged $20,000$ times.
A typical visibility matrix is shown here:
\begin{equation}
V = \left [\begin{array}{cccc}
0.959 & 0.015 & 0.025 & <0.001\\
 0.031 & 0.946 & 0.001 & 0.023 \\
 0.033 & 0.001 & 0.949 & 0.018 \\
 0.001 & 0.036 & 0.031 & 0.932 \\
\end{array}\right ]
\end{equation}
An ideal $V$ matrix is the identity matrix, with a trace equal to $4$.
In this example, the trace is $3.79$, which is equivalent to a total visibility of $\SI{94.65}{\percent}$ for the multiplexed readout.
Through linear inversion of this matrix we apply a readout correction to our measurement, giving $P_\mathrm{exp}$ in terms of $P_\mathrm{meas}$.
We note that this correction should not affect the quantum process fidelities calculated in the main text, which we have checked separately.

By contrast, for the Bell state fidelity, this readout correction does affect the result, and the results reported in the main text are not corrected in this way, as stated there.

For completeness, we note that using the readout correction increases the state fidelity from $\SI{72}{\percent}$ to $\SI{79}{\percent}$.
Finally,  the qubit excited state populations reported in the main text are calculated using $P_e^{Q_1} = P_{|eg\rangle}+P_{|ee\rangle}$ and  $P_e^{Q_2}=P_{|ge\rangle}+P_{|ee\rangle}$.

\section{Device parameters}

We summarize the device parameters in Table.~\ref{tab:supp-1}.

\begin{table*}[ht]
\begin{tabular}{lll}
\hline
\hline
Purcell filter & \multicolumn{2}{l}{} \\
\hline
Idle frequency [$\SI{}{\giga\hertz}$] & \multicolumn{2}{l}{$5.118$} \\
Coupling quality factor & \multicolumn{2}{l}{10} \\
\hline
\hline
Readout resonators & $R_1$ & $R_2$ \\
\hline
Idle frequency [$\SI{}{\giga\hertz}$] & $5.364$ & $5.318$ \\
Internal quality factor ($\times 10^3$)& $84$ & $86$ \\
Coupling quality factor ($\times 10^4$)& $4$ & $9$ \\
\hline
\hline
Qubits & $Q_1$ & $Q_2$ \\
\hline
Maximum frequency $f\left( \Phi=0\Phi_0 \right)$ [$\SI{}{\giga\hertz}$] & 5.03	 & 5.18 \\
Idle frequency [$\SI{}{\giga\hertz}$] & $4.221$ & $4.359$ \\
Anharmonicity [$\SI{}{\mega\hertz}$] & $198$ & $198$ \\
Thermal population without phononic thermalization & $0.0137$ & $0.0698$ \\
Thermal population with phononic thermalisation & $0.0047$ & $0.0021$ \\
Visibility  &\SI{94}{\percent} & \SI{94}{\percent} \\
Qubit capacitance (design value) [\SI{}{\femto\farad}] & $90$ & $90$ \\
Qubit SQUID inductance [\SI{}{\nano\henry}] & $10.8$ & $10.1$ \\
\hline
\hline
Qubit coherence at idle frequency &  & \\
\hline
Lifetime, $T_1$ [$\SI{}{\micro\second}$] & $57 \pm 15$ & $38 \pm 5$ \\
Ramsey dephasing time, $T_2^\mathrm{R}$ [$\SI{}{\micro\second}$] & $1.11 \pm 0.06$ & $0.88 \pm 0.04$ \\
Spin-Echo dephasing time, $T_2^\mathrm{e}$ [$\SI{}{\micro\second}$] & $3.8 \pm 0.1$ & $3.3 \pm 0.1$ \\
\hline
\hline
Qubit coherence at bidirectional frequency, $\SI{4.102}{\giga\hertz}$ &  & \\
\hline
Lifetime, $T_1$ [$\SI{}{\micro\second}$] & $38 \pm 15$ & $31 \pm 9$ \\
Ramsey dephasing time, $T_2^\mathrm{R}$ [$\SI{}{\micro\second}$] & $0.95 \pm 0.08$ & $0.62 \pm 0.06$ \\
Spin-Echo dephasing time, $T_2^\mathrm{e}$ [$\SI{}{\micro\second}$] & $2.48 \pm 0.06$ & $1.68 \pm 0.16$ \\
\hline
\hline
Qubit coherence at unidirectional transfer frequency, $\SI{3.976}{\giga\hertz}$ &  & \\
\hline
Lifetime, $T_1$ [$\SI{}{\micro\second}$] & $51 \pm 9$ & $33 \pm 5$ \\
Ramsey dephasing time, $T_2^\mathrm{R}$ [$\SI{}{\micro\second}$] & $0.79 \pm 0.08$ & $0.55 \pm 0.26$ \\
Spin-Echo dephasing time, $T_2^\mathrm{e}$ [$\SI{}{\micro\second}$] & $2.48 \pm 0.1$ & $2.26 \pm 0.28$ \\
\hline
\hline
Couplers & $G_1$ & $G_2$ \\
\hline
Coupler inductance [\SI{}{\pico\henry}] & $910$ & $873$ \\
Flip chip mutual (design) [\SI{}{\pico\henry}] & $160$ & $160$ \\
Grounding inductance (design) [\SI{}{\pico\henry}] & $480$ & $480$ \\
\hline
\hline
UDT & IDT & Mirror \\
\hline
Number of cells (design) & $20$ & $488$ \\
Aperture [\SI{}{\micro\meter}] (design) & $150$ & $150$ \\
Metallization ratio (SEM) & $0.52$ & $0.79$ \\
Reflectivity & $0.009$ & $-0.60 i$ \\
Wavelength [\SI{}{\micro\meter}] (design) & $0.975$ & $1$ \\
$\Delta v/v$ & \SI{3.44}{\percent} & \SI{2.70}{\percent} \\
Effective IDT-mirror distance  & \SI{140}{\nano\meter} & \\
Free space losses & \SI{173}{\neper\per\meter}& \\
\hline
\hline
\end{tabular}
\caption{Summary of device parameters}
\label{tab:supp-1}
\end{table*}

\section{Experimental setup}

\begin{figure*}
\centering
\includegraphics[width=0.9\textwidth]{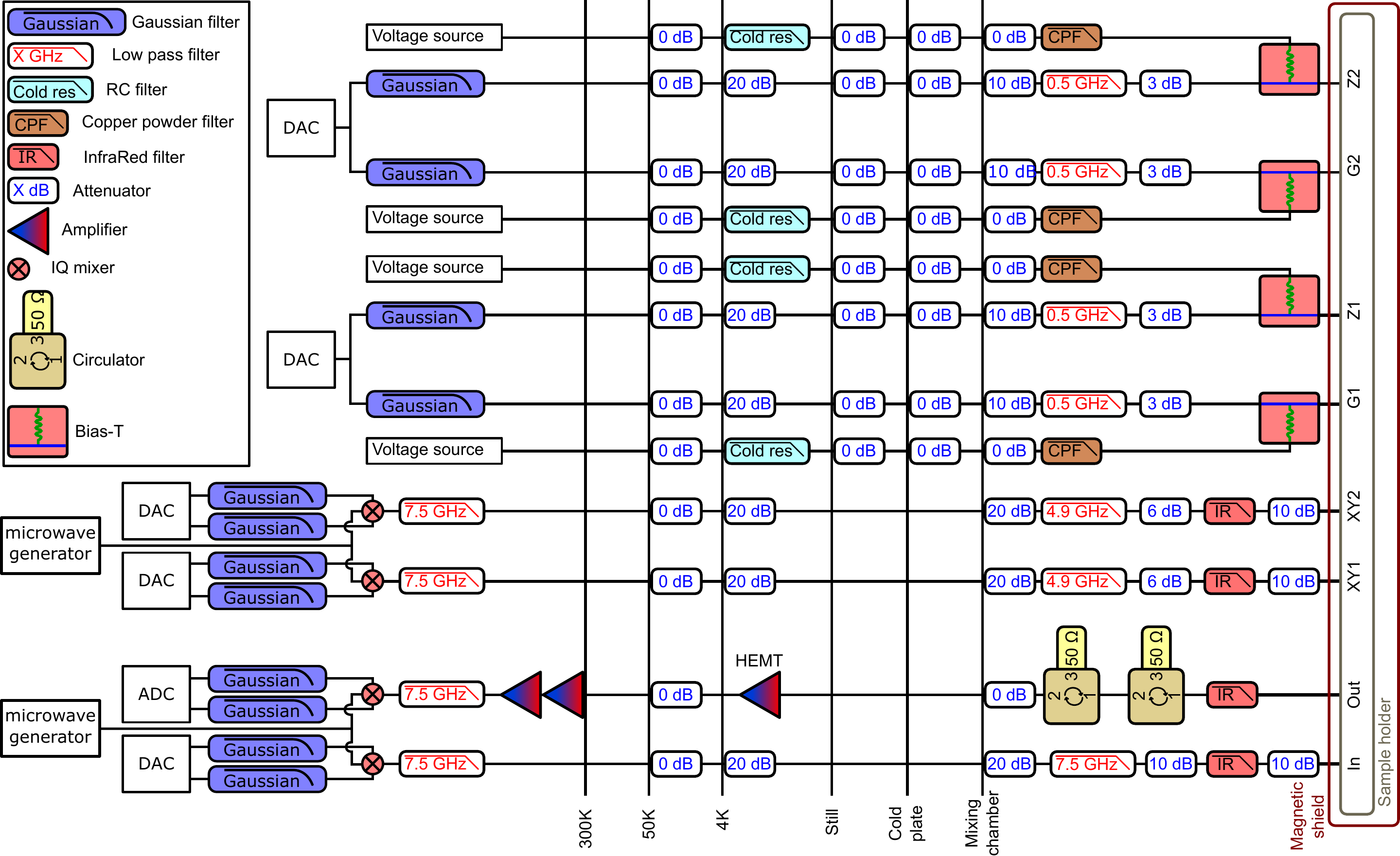}
\caption{Electronics and wiring diagram of the experimental setup.}
\label{fig.supp-wiring}
\end{figure*}

The assembled flip-chip, shown in the main text in Fig.~1, is wire bonded with 25 $\mu$m diameter aluminum wirebonds to an aluminum sample holder with copper traces.
The sample holder is placed inside a cryogenic magnetic shield and connected to the rest of the wiring using non-magnetic coaxial cables.
The full wiring and electronics are shown in Fig.~\ref{fig.supp-wiring}.
All experimental measurements are performed using a standard commercial dry dilution refrigerator with a base temperature below \SI{7}{\milli\kelvin}.

\section{Estimation of the dispersive shift}

The dispersive shift happening between the itinerant phonon and the qubit is estimated through a circuit model taking into account the full circuit i.e. the qubit, the coupler, the flip-chip connection and the UDT.
At this point, all circuit parameters have been set from design or obtained through fitting of other experiments.
From the interaction duration $\Delta t = \SI{200}{\nano\second}$ for Fig.~4\textbf{a} and $\Delta t = \SI{190}{\nano\second}$ for Fig.~4\textbf{b} and the phonon frequency $\omega_\mathrm{phonon}/(2\pi) = \SI{3.976}{\giga\hertz}$ we get a dispersive shift of
$\chi^\mathrm{exp} = \SI{1.00}{\mega\hertz}$ for Fig.~4\textbf{a} and $\chi^\mathrm{exp} = \SI{2.63}{\mega\hertz}$ for Fig.~4\textbf{b}.
From our model we estimate that during the interaction $Q_2$ frequency $\omega_{Q_2}/(2\pi)=\SI{4.190}{\giga\hertz}$ and its decay rate into the phononic channel $\kappa_{Q_2}/(2 \pi) = \SI{6}{\mega\hertz}$ for Fig.~4\textbf{a} and $\kappa_{Q_2}/(2 \pi) = \SI{7.65}{\mega\hertz}$ for Fig.~4\textbf{b}.
Approximating the UDT response to a harmonic oscillator, we obtain from the same circuit model a phononic emission rate $\kappa_\mathrm{UDT}/(2\pi)=\SI{147}{\mega\hertz}$, yielding through a resonant Purcell effect \cite{EMPurcell1946} to an effective qubit-itinerant phonon coupling $g/(2\pi)=\SI{14.85}{\mega\hertz}$ Fig.~4\textbf{a} and $g/(2\pi)=\SI{23.72}{\mega\hertz}$ for Fig.~4\textbf{b}.
From this we obtain a dispersive shift of $\chi^\mathrm{sim} = \SI{1.03}{\mega\hertz}$ for Fig.~4\textbf{a} and $\chi^\mathrm{sim} = \SI{2.63}{\mega\hertz}$ for Fig.~4\textbf{b}.

\end{document}